\documentclass{aa}

\usepackage{graphicx,natbib}
\bibpunct{(}{)}{;}{a}{}{,}

\def\LB{$\lambda$\,Bootis}
\def\Vmicro{$v_{\rm mic}$} 
\def\Vsini{$v\cdot\sin{i}$}
\def\Teff{$T_{\rm eff}$} 
\def\logg{$\log{g}$}

\def\trel{$t_{\rm rel}$}
\def\logP{$\log{P}$}
\def\kms{km\,s$^{-1}$}
\def\deg{$^{\circ}$}

\defcitealias{Heit:01a}{Paper~I}

\begin{document}

   \title{The accretion/diffusion theory for \LB\ stars\\
           in the light of spectroscopic data.
         }

   \author{U. Heiter\inst{1,2}
          \and
          W.W. Weiss\inst{1}
          \and
          E. Paunzen\inst{1,3}
          }

   \offprints{U. Heiter}

   \institute{Institute for Astronomy (IfA), University of Vienna,
              T\"urkenschanzstrasse 17, A-1180 Vienna, Austria\\
              email: weiss@astro.univie.ac.at
	 \and
	      Department of Astronomy, Case Western Reserve University,\\
	      10900 Euclid Avenue, Cleveland, OH 44106-7215, USA\\
	      email: ulrike@fafnir.astr.cwru.edu
         \and
              Zentraler Informatikdienst der Universit\"at Wien, 
	      Universit\"atsstrasse 7, A-1010 Vienna, Austria\\
              email: ernst.paunzen@univie.ac.at
             }

   \date{Received ; accepted }

\abstract{
Most of the current theories suggest the \LB\ phenomenon to originate from an interaction
between the stellar surface and its local environment.
In this paper, we compare the abundance pattern of the \LB\ stars to that of the
interstellar medium and find larger deficiencies for Mg, Si, Mn and Zn than in 
the interstellar medium. 
A comparison with metal poor post-AGB stars showing evidence for circumstellar material
indicates a similar physical process possibly being at work for some of the \LB\ stars,
but not for all of them.
Despite the fact that the number
of spectroscopically analysed \LB\ stars has considerably increased in the past,
a test of predicted effects with observations shows current abundance 
and temperature data to be still controversial.
}

   \maketitle


\section{Introduction}

\begin{figure}
  \resizebox{\hsize}{!}{\includegraphics[bb= 50 50 310 302]{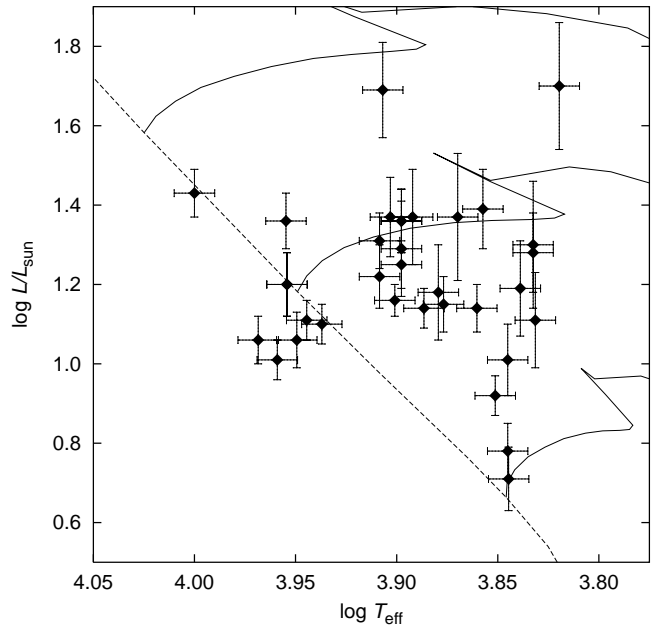}}
  \caption{Theoretical HR diagram for all \LB\ stars with known abundances.
           The luminosities have been derived from Hipparcos measurements, if
	   available, and from Str\"omgren photometry otherwise, as described
	   in \citet{Paun:00}, using data from \citet{Hauc:90}, \citet{Hand:95} and
	   \citet{Mart:98}. Evolutionary tracks for 1.5~M$_{\odot}$, 2~M$_{\odot}$ and 
	   2.5~M$_{\odot}$ taken from \citet{Clar:95} are also shown.}
  \label{evol}
\end{figure}

In the last few years, the \LB\ stars (metal-poor population~{\sc i}
A to F type stars) have experienced increased attention by abundance
analysis groups. The results have been collected by 
\citet[ hereafter referred to as Paper~I]{Heit:01a} and show that the proportion of \LB\ stars
with known abundances is now large enough to examine 
the abundances with respect to other stellar parameters on a good
statistical basis. The analysed
stars span a wide range of atmospheric parameters, in particular the
effective temperature (Fig.~\ref{evol}).
This parameter plays a major role in current theories,
which are briefly reviewed in the following.

\citet{Venn:90} proposed that the peculiar abundance patterns
of \LB\ stars originate from the interstellar medium, which shows a similar
abundance distribution (see Sect.~\ref{ISM}). Within this hypothesis it is 
assumed that only the interstellar gas, but not the dust, 
is accreted onto the surface of the stars. 
\citet{Char:91a} calculated the concentration of the elements Ca, Ti, Mn and Eu
in the superficial convection zone (SCZ) within a simple analytical model, which takes into account
accretion of interstellar gas and diffusion below the SCZ, for various effective
temperatures. It is assumed that
the atmospheric material is mixed thoroughly from the surface to the bottom of the SCZ.

More detailed numerical calculations have been performed by \citet{Turc:93}
for the elements Ca, Sc and Ti. Abundance profiles of the first two elements 
show overabundances at the surface if only chemical separation and convective 
mixing is taken into account, whereas Ti is predicted to be underabundant in 
this case. If accretion of circumstellar gas with a certain amount of depletion
is added, the calculations show that for an accretion rate $\dot M$ of at least
$5\cdot 10^{-14}$~M$_{\odot}$~yr$^{-1}$, the abundances of the examined
elements in the convection zone converge to the values in the accreted gas 
on a very short timescale. Two other points became evident. Independently
of $\dot M$ and the duration of the accretion phase, the abundances are
again governed by chemical separation when accretion is stopped.
This means that accretion must be an ongoing process if it is responsible
for the observed abundance pattern. Secondly, meridional circulation induced
by rotation with equatorial velocities up to 125~\kms\ does not alter
the surface abundances produced by accretion. Higher rotation rates could
not be treated due to numerical problems. All these calculations are based
on only one static stellar envelope model with a fixed parameter set
of (\Teff, \logg) = (8000~K, 4.3).

The separation of gas and dust in a circumstellar shell was investigated
by \citet{Andr:00}, who calculated gas and dust grain velocities in a shell extending
to 100 stellar radii around a star with \Teff=8500~K, assuming a polytropic
density distribution. For a ratio between radiative and
gravitational acceleration on the gas of 0.99, large dust grains and a rather smooth density
distribution (polytropic index =~2), they indeed find dust grains to be
forced to an outward motion by radiative pressure. The separation becomes effective at
a distance from the stellar
surface where the temperature is about 1600~K (condensation temperature for heavy elements),
which corresponds to about 10 stellar radii.
The gaseous part of the shell is accreted to the surface of the star.
Thus the two components are decoupled and the superficial chemical composition
is changed according to the depletions in the gas coming from the outer part
of the shell. The calculations take into account only interactions between neutral
particles because they are shown to be more important than Coulomb-type interactions.
Within this simple model only rough estimates for the gas-dust separation can be made,
which are based on very restricted assumptions. But it could serve for more
sophisticated models, which in particular should be extended to lower temperatures and
smaller dust grains, which are more likely to be formed around \LB\ stars.

An earlier theoretical approach to the \LB\ phenomenon had its origin
in recalling the physical processes operating in the atmospheres of Am stars.
The Am abundance pattern has been explained by \citet{Char:91b} by chemical separation
of elements below the superficial hydrogen convection zone, 
caused by diffusion processes.
In order to produce the
actual abundance values of Am stars, an additional process is needed,
e.g. a small amount of mass-loss ($10^{-15}$~M$_{\odot}$~yr$^{-1}$).
Evidence for mass-loss has not yet been observed for Am stars.
By introducing a two orders of magnitude higher mass-loss rate,
\citet{Mich:86} have changed the calculated Am abundance pattern to a \LB\ like
one, but they have not been able to produce underabundances as low
as were observed in 
several \LB\ stars. The underabundances even vanish for most elements if
meridional circulation induced by high rotational velocities
is taken into account \citep{Char:93}.
Although large uncertainties are still involved
in the modeling (above all for the radiative acceleration), 
this theory has been widely discarded as an explanation of the
\LB\ star abundances.

Further theoretical considerations include \citet{Andr:97}, who proposed that
\LB\ stars are the result of a merger of contact binaries of W~UMa type.
He argues that mass loss during the merger phase could form the circumstellar
shell, whose accretion leads to the observed underabundances.
The hypothesis is substantiated by lifetime and number estimates.
\citet{Fara:99} suspect that a part of the \LB\ stars are
undetected spectroscopic binary systems, and that their abundance anomalies
are due to veiling effects in the composite spectra.

Summarizing, in the mentioned theories the \LB\ phenomenon seems to originate from an interaction
between the stellar surface and its local environment. In the following we 
confront predicted effects with observations.

\section{Examination of stellar parameters and abundances}
\label{correlations}

\subsection{Correlations between stellar parameters}

First, we searched for correlations among the atmospheric parameters \Teff\ and \logg, 
the projected rotational velocity, the relative age (\trel) and
the pulsational period, all listed in Table~\ref{par_all_tab}.
A correlation appeared to exist between \Teff\ and \logg\ and also
between \trel\ -- \Teff\ and \trel\ -- \logg. In all three cases the errors of the
slopes were smaller than 20~\%. However, these correlations seem to be due to a
selection effect, which
becomes clear from Fig.~\ref{evol}, showing
the location of the analysed stars in the HR diagram, as well as some evolutionary
tracks from \citet{Clar:95}. The lack of young stars in cooler regions and old stars
in hotter regions is evident. This imbalance disappears when the same diagram is regarded
for all \LB\ stars \citep[ Fig.~33]{Paun:00}.

The \Teff~-~\Vsini\ relationship basically follows that of normal dwarfs
\citep{Schm:82}, which shows that a high \Vsini\
value cannot be used as a \LB\ classification criterion.
No correlation exists between \Vsini\ and \trel.

\begin{table*}
\caption{Stellar parameters of the \LB\ stars for which abundance analyses have been performed up to now.
The atmospheric parameters have been taken from the sources listed in Table~6 of 
\citetalias{Heit:01a}.
The column labeled $\sigma$[X] lists the standard deviations of 7-9 heavy element
abundances as described in the text.
The age \trel\ is given relative to the main sequence lifetime
and has been estimated, like the mass, using the CESAM code \citep{More:97}.
The pulsational parameters are from \citet{Paun:97d,Paun:98a}. 
Column $A/UL$ gives amplitudes for variable and upper limits for constant stars
with the used filter in parentheses.
The coordinates have been extracted from the Simbad database.
The distances $r$ have been calculated using Hipparcos parallaxes \citep{Perr:97}.
The errors for \Vsini and $r$ are given in parentheses.}
\label{par_all_tab}
\begin{tabular}{@{}rrl@{\hspace*{2mm}}rr@{}rrrrrr@{}lrrr@{}r}
\hline\hline
   & \Teff & \logg & \Vmicro & \multicolumn{2}{c}{\Vsini} & $\sigma$[X] & \trel & $M$ & \logP & $A/UL$ &       & $l$ & $b$ & \multicolumn{2}{c}{$r$} \\
   & $\pm$200 & $\pm$0.3 & $\pm$0.5 &       & & & $\pm$1 & $\pm$0.05 &           & &        &        &     & \\
HD & [K] & \multicolumn{4}{l}{[log cm s$^{-1}$] [\kms]} & & [\%] & [M$_{\odot}$] & [log~d] & [mmag] & & [\deg] & [\deg] & \multicolumn{2}{c}{[pc]} \\
\hline
   319 & 8100 & 3.8 & 3.5 &  60 & (5) & 0.25 &  66 & 2.1 &         &  4.2 & (b) &  55.56 & $-$79.07 &  80 &  (5) \\
 11413 & 7900 & 3.8 & 3.0 & 125 &(10) & 0.41 &  36 & 1.6 & $-$1.43 & 18.0 & (B$-$V) & 280.70 & $-$64.28 & 75 &  (4) \\
 15165 & 7200 & 3.7 & 3.0 &  90 &(10) & 0.39 & 100 & 2.0 & $-$0.90 & 20.1 & (V) & 157.65 & $-$45.78 & 118 & (14) \\
 31295 & 8800 & 4.2 & 3.0 & 110 &(10) & 0.29 &   9 & 2.0 &         &  7.4 & (v) & 189.34 & $-$20.25 &  37 &  (1) \\
 74873 & 8900 & 4.6 & 3.0 & 130 &(10) & 0.25 &   3 & 2.0 &         &  9.4 & (v) & 214.84 &    31.05 &  61 &  (5) \\
 75654 & 7250 & 3.8 & 3.0 &  44 & (5) &      &  87 & 1.8 & $-$1.06 & 20.0 & (V) & 260.52 &     3.08 &  78 &  (4) \\
 81290 & 6780 & 3.5 & 3.0 &  56 & (5) &      & 100 & 1.8 &         &  2.4 & (b) & 271.85 &     0.80 &     &      \\
 83041 & 6900 & 3.5 & 3.0 &  95 &(10) &      & 100 & 1.9 & $-$1.18 &  7.0 & (b) & 259.25 &    16.88 &     &      \\
 84123 & 6800 & 3.5 & 3.0 &  15 & (2) & 0.18 & 100 & 1.9 &         &  3.0 & (b) & 178.99 &    49.25 & 110 & (12) \\
84948A & 6600 & 3.3 & 3.5 &  45 & (5) & 0.21 & 100 & 2.3 &         &      &   & 168.14 &    48.97 & 201 & (60) \\
84948B & 6800 & 3.7 & 3.5 &  55 & (5) & 0.15 & 100 & 1.9 & $-$1.11 & 15.0 & (v) & 168.14 &    48.97 & 201 & (60) \\
101108 & 7900 & 4.1 & 3.0 &  90 &(10) & 0.15 &  66 & 1.8 &         &  2.0 & (b) & 171.03 &    70.82 & 233 &(155) \\
105759 & 8000 & 4.0 & 3.0 & 120 & (5) &      &  86 & 2.1 & $-$1.17 & 11.9 & (B) & 285.61 &    53.71 &     &      \\
106223 & 7000 & 4.3 & 3.0 & 100 &(10) & 0.20 & 100 & 1.7 &         &  3.0 & (b) & 190.02 &    81.06 & 110 & (11) \\
107233 & 7000 & 3.8 & 3.0 &  95 &(15) & 0.28 &  21 & 1.6 &         &      &   & 297.54 &    14.22 &  81 &  (6) \\
109738 & 7575 & 3.9 & 3.0 & 166 &(10) &      &  79 & 1.9 & $-$1.49 & 18.0 & (b) & 301.63 &  $-$5.03 &     &      \\
110411 & 9100 & 4.5 & 3.0 & 160 &(10) & 0.31 &   0 & 2.0 &         &      &   & 294.88 &    72.96 &  37 &  (1) \\
111005 & 7410 & 3.8 & 3.0 & 138 &(10) &      & 100 & 2.1 &         &      &   & 300.10 &    65.26 & 174 & (37) \\
125162 & 8650 & 4.0 & 3.0 & 100 &(10) & 0.59 &  19 & 2.0 & $-$1.64 &  2.0 & (b) &  86.97 &    64.67 &  30 &  (1) \\
142703 & 7100 & 3.9 & 3.0 &  95 &(10) & 0.22 &  52 & 1.6 & $-$1.38 &  6.0 & (b) & 355.62 &    28.55 &  53 &  (2) \\
156954 & 6990 & 4.1 & 3.0 &  51 & (5) &      &  25 & 1.5 &         &  2.6 & (b) &  10.50 &    13.19 &  82 &  (7) \\
168740 & 7700 & 3.7 & 3.0 & 130 &(10) & 0.21 &  74 & 1.9 & $-$1.44 & 16.0 & (b) & 331.84 & $-$21.15 &  71 &  (4) \\
170680 & 10000 & 4.1 & 2.0 & 200 &(10) &     &  40 & 2.4 &        &      &   &  14.30 &  $-$4.03 &  65 &  (4) \\
171948A & 9000 & 4.0 & 2.0 & 15 & (2) &      &  15 & 2.0 &         &  2.6 & (b) &  51.34 &    13.00 & 131 & (14) \\
171948B & 9000 & 4.0 & 2.0 & 10 & (2) &      &  15 & 2.0 &         &  2.6 & (b) &  51.34 &    13.00 & 131 & (14) \\
183324 & 9300 & 4.3 & 3.0 &  90 &(10) & 0.31 &   0 & 2.0 & $-$1.68 &  4.0 & (v) &  38.99 &  $-$7.45 &  59 &  (3) \\
192640 & 7960 & 4.0 & 3.0 &  80 & (2) & 0.25 &  60 & 1.9 & $-$1.52 & 26.0 & (b) &  74.45 &     1.17 &  41 &  (1) \\
193256 & 7800 & 3.7 & 3.0 & 250 &(25) & 0.43 & 100 & 2.2 &         &  2.6 & (b) &  13.61 & $-$31.10 & 218 &(116) \\
193281 & 8070 & 3.6 & 2.8 &  97 &(10) &      & 100 & 2.6 &         &  3.4 & (b) &  13.60 & $-$31.11 & 218 &(116) \\
198160 & 7900 & 4.0 & 3.0 & 200 &(20) & 0.46 &  85 & 2.1 &         &      &   & 333.32 & $-$37.62 &  73 &  (7) \\
198161 & 7900 & 4.0 & 3.0 & 180 &(20) & 0.44 &  85 & 2.1 &         &      &   & 333.32 & $-$37.62 &  73 &  (7) \\
204041 & 8100 & 4.1 & 3.0 &  65 &(10) & 0.39 &  59 & 1.9 &         &  1.8 & (b) &  53.47 & $-$33.44 &  87 &  (8) \\
210111 & 7530 & 3.8 & 2.9 &  56 & (5) & 0.56 &  79 & 1.8 & $-$1.33 &  8.0 & (V) &  13.11 & $-$54.53 &  79 &  (6) \\
221756 & 9010 & 4.0 & 3.0 & 100 &(10) & 0.18 &  66 & 2.2 & $-$1.36 &  7.0 & (b) & 107.40 & $-$20.31 &  72 &  (3) \\
\hline\hline
\end{tabular}
\end{table*}

\subsection{Correlations between stellar parameters and abundances}
\begin{figure*}
  \resizebox{\hsize}{!}{\includegraphics{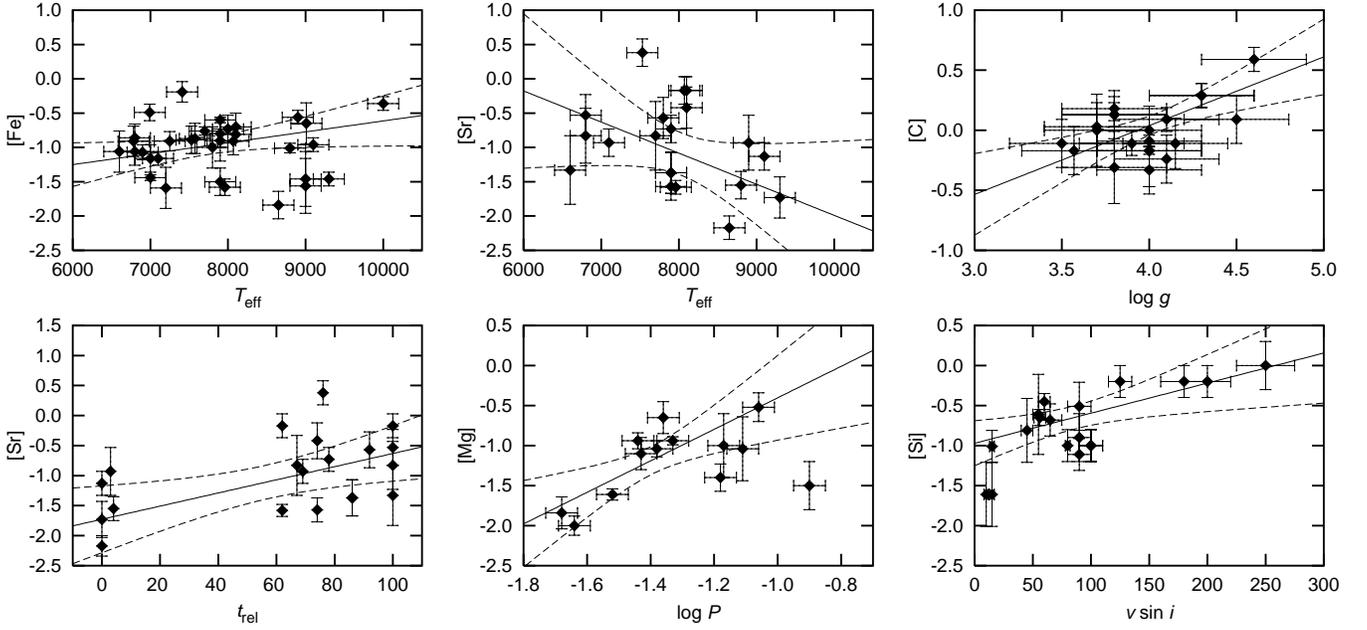}}
  \caption{Dependences of element abundances on the different parameters.
           This figure contains only ``best cases'', to avoid overloading this
	   paper with graphs. The dashed lines indicate the 95\% confidence bands 
           of the weighted linear fit (solid line).}
  \label{corr_fig}
\end{figure*}

\begin{figure*}
  \resizebox{\hsize}{!}{\includegraphics[bb=50 215 410 302]{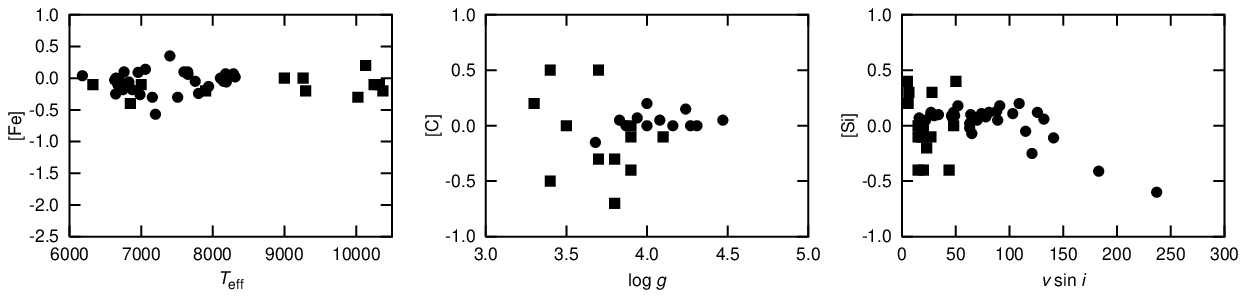}}
  \caption{Distribution of Fe, C and Si abundances with \Teff, \logg\ and \Vsini, respectively, for normal A and F stars in the Hyades (circles) and in the field (squares).}
  \label{corr_fig_n}
\end{figure*}

As a next step, a linear regression, weighted by $1/s^2$ according to the 
errors of the abundances 
($s$, taken from the references given in Table~6 of \citetalias{Heit:01a}),
has been computed for each element with respect to each
of the five parameters discussed in the previous subsection. 
For most elements the errors of the slopes appeared to be very large compared to their values.
For some elements with an error between 55 and 100\% 
a very weak correlation might be indicated.
These cases are marked in Table~\ref{corr_tab} with one plus sign (positive slope) or
one minus sign (negative slope).
For Cr, Fe and Sr vs. \Teff, for Sc vs. \trel, for Ca vs. \logP\ and for Mg vs. \Vsini, 
the errors of the slopes amount to about 50\%, and they lie between 30 and 40\%
for Sc vs. \Teff, C and O vs. \logg, Na and Sr vs. \trel, Mg vs. \logP\ and
Si and Fe vs. \Vsini.
These cases may indicate a weak correlation and are represented by two plus or
minus signs in Table~\ref{corr_tab}. Some of them are shown in Fig.~\ref{corr_fig},
together with 95\% confidence bands of the fits.
However, within a 90\% confidence level the hypothesis of no correlation
between parameters and abundances is possible for {\em all} elements and parameters.
Plots for the same elements and parameters for normal stars, where available
(Fig.~\ref{corr_fig_n}, see \citetalias{Heit:01a} for the references), 
do not show any correlation except maybe for [Si] -- \Vsini.

\begin{table}
\caption{Formal correlations between stellar parameters and abundances. Double signs indicate
a weak correlation (error of the slope lower than 55\%), single signs indicate a
very weak correlation (error of the slope between 55 and 100\%), whereas all other
quantities are clearly uncorrelated. + and $-$ signs correspond to positive and
negative slopes.}
\label{corr_tab}
\begin{tabular}{lrrrrr}
\hline\hline
El.   & \Teff      & \logg  & \trel & \logP & \Vsini \\
\hline
C     &            & ++     & $-$   &       &        \\
C$_{\rm NIR}$ & +  &  +     &       &       & +      \\
O     &            & $--$   &       &       &        \\
O$_{\rm NIR}$ & +  &        & $-$   &       & $-$    \\
Na    &            & $-$    & ++    &       &        \\
Mg    &            & $-$    & +     & ++    & ++     \\
Si    & $-$        &        & +     &       & ++     \\
Ca    &            &        & +     & ++    &        \\
Sc    & ++         &        & $--$  & $-$   &        \\
Ti    & +          &        & +     & +     & +      \\
Cr    & ++         &        &       &       &        \\
Fe    & ++         & +      &       & +     & ++     \\
Ni    & +          & +      &       &       &        \\
Sr    & $--$       & $-$    & ++    &       & $-$    \\
Ba    &            &        &       &       & $-$    \\
\hline\hline
\end{tabular}
\end{table}

\subsection{The effective temperature}
\label{teff}
Within the accretion/diffusion model, the abundances depend mainly on 
the $\dot M$, the mass of the convection zone, and the radiative acceleration.
One prediction of this model is that underabundances
are most likely to appear in a certain temperature interval, in which accretion
dominates the diffusion processes. This interval begins at 7000~K for all elements,
and its width depends on the element and $\dot M$.
For Ti, it extends at least to 9500~K, for Mn at least to 8000~K, and for Ca well over
12000~K. For higher temperatures, the abundance effects depend more on $\dot M$ 
than for lower temperatures.
Observational implications are \citep{Char:91a}:\\
{\em Cool border:} 7000~K for all accretion rates and elements.\\
{\em Hot border:} Abundance variations from element to element
and from star to star should be more pronounced among the hotter \LB\ stars than among the cooler ones.

The most recent list of classified \LB\ stars can be found in
\citet{Paun:00}. The temperatures of 
the stars subject to abundance analyses (Table~\ref{par_all_tab}) span the
whole temperature range given by the stars in this list.
In the temperature range of 6500 to 8500~K, 45\% of the 56 \LB\ stars, 
and in the range of 8500 to 10500~K, about 55\% of the 16 \LB\ stars 
have been analysed.
The theoretically predicted {\em cool border} of the \LB\ group
is not supported by our observations because the coolest \LB\ star in our sample has a
\Teff\ of 6600~K. 
However, the treatment of convection used in the accretion/diffusion model plays a
crucial role for modeling these borders.
For example, increasing $\alpha$ 
(the ratio of mixing length to pressure scale height) from 1.4 to 2.0 would shift the
cool border for Ti by 400~K towards higher values \citep{Char:91a}.
Similarly, a decrease of $\alpha$ (values of 1.25 down to 0.5 are commonly used 
in models of main sequence stars) or a different convection model 
could result in a shift towards lower temperatures. 

For hot stars, the accretion/diffusion theory predicts
different degrees of heavy element depletion.
For an accretion rate of $5\cdot 10^{-14}$~M$_{\odot}$~yr$^{-1}$,
Ca should still be underabundant for a \LB\ star above 9000~K, 
while Ti can be solar abundant. This difference vanishes for 
$\dot M \ge 2\cdot 10^{-13}$~M$_{\odot}$~yr$^{-1}$ \citep[ Fig.~1]{Char:91a}.
In all five stars with \Teff\ greater than about 9000~K and abundances
determined for Ca and Ti (this includes HD\,31295), these elements have the same abundance.
At the cool border, regardless of $\dot M$, all elements should 
become solar abundant at the same temperature.
To test these predictions (arbitrarily extended to all elements) 
we calculated the standard deviations of the heavy element abundances 
(relative to solar) for individual stars, where at least seven 
of the elements Mg, Si, Ca, Sc, Ti, Cr, Fe, Sr and Ba have been measured
(see the column labelled $\sigma$[X] in Table~\ref{par_all_tab}). 
We did not find any dependence of the abundance scatter on effective temperature.

A positive trend of abundances with temperature is indicated for Sc, Cr and Fe
(Table~\ref{corr_tab}).
In the context of Charbonneau's model, this would imply a decrease of
$\dot M$ with increasing \Teff. On the other hand,
Sr shows a negative trend and Na behaves totally differently, 
because pronounced overabundances occur around 8000~K and they 
decrease towards both temperature borders.

For underabundances, the accretion/diffusion theory predicts a rather narrow temperature interval of
7000 $-$ 9000~K for Mn and 7000 $-$ 8000~K for Eu.
The mean Mn abundance of the five \LB\ stars with temperatures from 6800 to 7250~K 
is $-$1.2~$\pm$0.3~dex, whereas the abundances of the remaining three stars 
($\approx$~8000~K) span a large range ($-$0.4, $-$1.0 and $-$2.1~dex), which
might be caused by different accretion rates.
Due to the lack of Eu abundance measurements, the behavior of this element cannot be tested.
No accretion/diffusion calculations are available for the light elements,
whose nearly solar abundances are characteristic for \LB\ stars. 

The discussion above shows that current abundance and temperature data 
are inconclusive with regard to predictions of the accretion/diffusion theory.

\subsection{The other parameters}
The marginal C abundance correlation with
\logg\ (Fig.~\ref{corr_fig}) could indicate a decrease with progressing evolution.
On the other hand, Table~\ref{corr_tab} indicates an opposite trend for other elements. 
A connection to the stellar age is also present for the pulsational period
via the $Q$-constant,
which is the suspected reason for the positive slope present for the abundances 
of Mg, Ca, Ti and Fe with respect to \logP.

\citet{Char:91a} has calculated the critical
values for the equatorial rotational velocity, for various effective temperatures
and accretion rates, above which the rotational mixing is too strong for
a star to show abundance anomalies (see his Fig.~2). These can be interpreted as upper limits
for the \Vsini\ values of \LB\ stars. 
For Ti, the maximum velocities at 9000~K are 170~\kms\ ($5\cdot 10^{-14}$~M$_{\odot}$~yr$^{-1}$), 250~\kms\ ($10^{-13}$~M$_{\odot}$~yr$^{-1}$) and 370~\kms\ ($2\cdot 10^{-13}$~M$_{\odot}$~yr$^{-1}$).
The values of \Vsini\ and \Teff\ of the
\LB\ stars in our sample are consistent with
$\dot M = 10^{-13}$~M$_{\odot}$~yr$^{-1}$.
The abundances of the heavy elements Mg, Si, Ti
and Fe seem to increase with \Vsini\ (Fig.~\ref{corr_fig} for the best case of Si 
and Table~\ref{corr_tab}), which could
indicate that rotational mixing prevents the development of large
underabundances for higher rotational velocities.
Note that an opposite trend seems to exist for normal stars
(Fig.~\ref{corr_fig_n}).


\section{Comparison with the interstellar medium}
\label{ISM}

\begin{table*}
\caption{Characteristic parameters for the interstellar medium along the sight lines to several stars. The fifth column contains the total hydrogen column density, and the consecutive columns give for each element $X$ the value $\log\,\left(\frac{N(X)}{N(\rm H)}\right) - \log\,\left(\frac{N(X)}{N(\rm H)}\right)_{\odot}$. The unit of $N(X)$ is cm$^{-2}$. Column twelve gives the references. The asterisks denote sight lines with coordinates in the vicinity of \LB\ stars. Errors are given in parentheses.}
\label{ISM_tab}
\begin{tabular}{@{}rcrrr@{}r@{}rrrrrrrl@{}}
\hline\hline
(1) & & (2) & (3) & \multicolumn{2}{c}{(4)} & (5) & (6) & (7) & (8) & (9) & (10) & (11) & (12) \\
HD & & $l$ [\deg] & $b$ [\deg] & \multicolumn{2}{c}{$r$ [pc]} & $\log{N(\rm H)}$ & [O] & [Mg]     & [Si]        &  [S]        & [Fe]        & [Zn]    & Ref. \\
\hline
  5394 &   & 123.6 & $-$2.2  & 188  & (22)  & 20.18 & $-$0.3(1) &           &           &           &           &          & 1 \\
 18100 &   & 217.9 & $-$62.7 & 3100 &       & 20.14 &           & $-$0.8(1) & $-$0.4(1) & $-$0.3(1) & $-$0.8(1) & $-$0.2(2) & 2,3 \\
 22586 & * & 264.2 & $-$50.4 & 2000 &       & 20.35 &           &           & $-$0.6(2) & $-$0.6(2) & $-$1.2(2) &          & 2 \\
 24398 &   & 162.3 & $-$16.7 & 301  & (88)  & 21.20 & $-$0.4(1) &           &           &           &           &          & 1 \\
 24760 &   & 157.4 & $-$10.1 & 165  & (26)  & 20.52 & $-$0.4(1) &           &           &           &           &          & 1 \\
 24912 &   & 160.4 & $-$13.1 & 543  & (334) & 21.29 & $-$0.4(1) & $-$1.2(1) &           &           & $-$1.9(1) & $-$0.6(1) & 1,2,4 \\
 35149 & * & 199.2 & $-$17.9 & 295  & (102) & 20.74 &           &           &           &           &           & $-$0.3(3) & 2,4\\
 36486 & * & 203.9 & $-$17.7 & 281  & (85)  & 20.17 & $-$0.4(1) &           &           &           &           & $-$0.2(3) & 1,4 \\
 36861 & * & 195.1 & $-$12.0 & 324  & (109) & 20.81 & $-$0.4(1) &           &           &           &           &          & 1 \\
 37043 &   & 209.5 & $-$19.6 & 407  & (185) & 20.18 & $-$0.3(1) &           &           &           &           &          & 1 \\
 37128 & * & 205.2 & $-$17.2 & 412  & (246) & 20.48 & $-$0.4(1) &           &           &           &           & $-$0.2(3) & 1,4 \\
 38666 &   & 237.3 & $-$27.1 & 397  & (111) & 19.85 &           & $-$0.4(4) & $-$0.5(3) &    0.0(1) & $-$1.1(1) & $-$0.1(1) & 2 \\
 38771 &   & 214.5 & $-$18.5 & 221  & (45)  & 20.53 & $-$0.4(1) &           &           &           &           &          & 1 \\
 44743 &   & 226.0 & $-$14.0 & 153  & (17)  & 19.28 &           &           &    0.0(4) &           &           &          & 5 \\
 47839 &   & 202.9 & 2.2     & 313  & (93)  & 20.40 & $-$0.5(1) &           &           &           &           & $-$0.2(3) & 1,4 \\
 57061 &   & 238.2 & $-$5.5  & 1514 &       & 20.80 & $-$0.3(1) &           &           &           &           & $-$0.2(3) & 1,4 \\
 68273 &   & 262.8 & $-$7.7  & 258  & (41)  & 19.74 &           & $-$0.4(2) & $-$0.7(2) &           & $-$1.1(3) &          & 2 \\
 72089 &   & 263.2 & $-$3.9  & 1700 &       & 20.60$^{\dagger}$ & & $-$1.1(2) & $-$0.5(3) &         & $-$0.5(4) &          & 2 \\
 91316 &   & 234.9 & 52.8    & 869  &       & 20.44 &           &           &           &           &           & $-$0.2(3) & 4 \\
 93521 & * & 183.1 & 62.2    & 1700 &       & 20.10 &           &    0.0(4) & $-$0.4(2) & $-$0.1(1) & $-$0.7(4) &          & 2 \\
100340 &   & 258.8 & 61.2    & 5300 &       & 20.47 &           & $-$0.8(2) &           & $-$0.4(1) &           &          & 3 \\
116852 &   & 304.9 & $-$16.1 & 4800 &       & 20.96 &           & $-$0.8(2) &           & $-$0.3(2) & $-$0.8(2) & $-$0.3(2) & 2 \\
120086 &   & 329.6 & 57.5    & 299  & (140) & 20.41 &           & $-$0.9(2) & $-$0.7(2) & $-$0.7(2) & $-$1.3(2) &          & 2 \\
141637 & * & 346.1 & 21.7    & 160  & (27)  & 21.20 &           &           &           &           &           & $-$0.4(3) & 2,4 \\
143018 & * & 347.2 & 20.2    & 141  & (19)  & 20.75 &           &           &           &           &           & $-$0.3(3) & 2,4 \\
144217 & * & 353.2 & 23.6    & 163  & (36)  & 21.08 &           &           &           &           &           & $-$0.2(3) & 4 \\
147165 & * & 351.3 & 17.0    & 225  & (50)  & 21.40 &           &           &           &           &           & $-$0.4(3) & 4 \\
149757$^{\ddagger}$ & * & 6.3 & 23.6      & 140  & (16)  & 21.13 & $-$0.4(1) & $-$1.6(1) & $-$1.3(1) &    0.1(4) & $-$2.1(1) & $-$0.7(1) & 1,2,4 \\
                    &   &     &           &      &       & 19.74 &    0.0(3) & $-$0.9(1) & $-$0.5(1) &           & $-$1.1(1) &    0.0(1) & 2 \\
149881 &   & 31.4 & 36.2     & 2100 &       & 20.57 &           &    0.0(2) & $-$0.3(3) & $-$0.1(2) & $-$0.8(4) &    0.0(3) & 2,4 \\
154368 & * & 350.0 & 3.2     & 366  & (199) & 21.62 & $-$0.5(1) & $-$0.6(3) & $-$0.8(5) & $-$0.6(3) & $-$1.2(3) &    0.1(3) & 2 \\
157246 & * & 334.6 & $-$11.5  & 348 & (123) & 20.73 & $-$0.3(1) &           &           &           &           &          & 1 \\
158926 & * & 351.8 & $-$2.2   & 216 & (52)  & 19.23 &           &           &           &           &           & $-$0.1(3) & 4 \\
167756 &   & 351.5 & $-$12.3  & 4000 &      & 20.81 &           & $-$0.9(1) &           &           &           & $-$0.2(1) & 2,4 \\
212571 & * & 66.0 & $-$44.7  & 338  & (107) & 20.56 &           &           &           & $-$0.3(3) & $-$0.9(4) &    0.0(3) & 2,4 \\
215733 &   & 85.2 & $-$36.4  & 2900 &       & 20.76 &           & $-$0.2(3) & $-$0.6(2) &           &           & $-$0.3(2) & 4,6 \\
3C 273 & * & 290.0 & 64.4    &      &       & 20.10 &           &           & $-$0.7(2) &    0.1(2) &           &           & 2 \\
\hline\hline
\\
\multicolumn{14}{l}{References: (1) \citet{Meye:98} (2) references given in \citet[ their Table~3]{Sava:96a}} \\
\multicolumn{14}{l}{(3) \citet{Sava:96b} (4) \citet{Roth:95} (5) \citet{Dupi:98} (6) \citet{Fitz:97}} \\
\multicolumn{14}{l}{$^{\dagger}$ \citet{Dubn:98}, $^{\ddagger}$ cool (first row) and warm (second row) clouds} \\
\end{tabular}
\end{table*}

\begin{figure}
  \resizebox{\hsize}{!}{\includegraphics{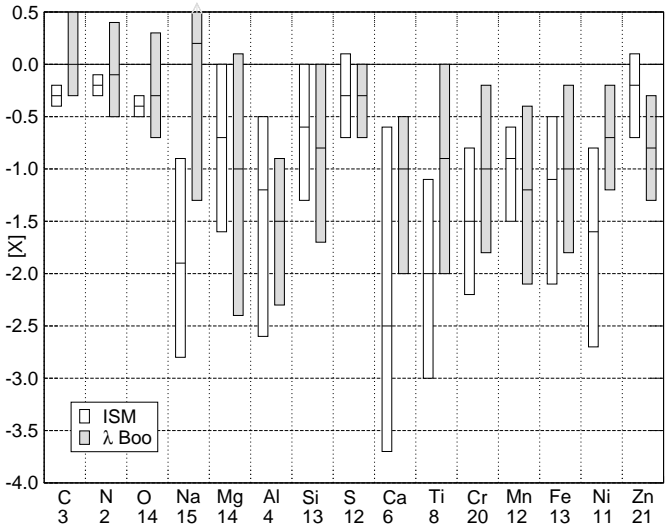}}
  \caption{Mean abundances for all ISM sight lines from Table~\ref{ISM_tab}, as well as highest and lowest abundances. The number of available abundances is given below the element name. The grey bars indicate the mean and ranges of \LB\ star abundances.}
  \label{ism_mean}
\end{figure}

The accretion hypothesis was proposed by \citet{Venn:90} because they
noticed similarities between \LB\ star underabundances and element depletions
in the interstellar gas. They assumed that
gas in a dusty circumstellar nebula, whose accretion could result in the observed
underabundances, shows depletions similar to those of the interstellar gas.
Therefore we compare in this Section the abundance pattern of the \LB\ stars
\citepalias{Heit:01a} with recent determinations of abundances in the interstellar gas.

The chemical composition of the interstellar gas has been studied
by several authors along many different sight lines.
\citet{Sava:96a} give an extensive review on abundances in the interstellar medium (ISM).
The galactic longitudes, latitudes and distances of the stars at the end of the sight lines
are given in columns two to four of Table~\ref{ISM_tab}.
The hydrogen column density is given in column five and the depletions of six elements 
are given in columns six to eleven. The solar values have been taken from
\citet{Grev:96}. 
Abundances for the following additional elements, which have also been studied in
\LB\ stars, are available: C, N, Na \citep{Welt:94}, Al, Ca, Ti, Cr, Mn, Co, Ni, Cu.
For Na, only absorption lines of the
neutral element have been studied, whereas singly ionized Na is more abundant 
in the ISM. Therefore these abundances cannot be compared
directly to that of the other elements, except for the sight line to
$\zeta$~Oph (HD\,149757). For this star, 
a ratio of the ionized to neutral Na column density of 3.1
was derived by \citet{Mort:75}.

\subsection{Mean abundances}
The depletions of the individual elements in the ISM have been averaged
over all sight lines and the results are displayed in Fig.~\ref{ism_mean},
as well as the highest and lowest measured relative abundances. 
For most elements the abundances vary considerably
for the different regions that have been analysed. 
For comparison, the \LB\ star abundance pattern \citepalias{Heit:01a} 
is also shown in Fig.~\ref{ism_mean} (grey bars).
The scatter of element abundances within the two samples is similar,
except for O, which seems to be distributed very homogeneously throughout the ISM. 
Another similarity between the \LB\ group and the ISM are the mean abundances of 
C and O, which are slightly below solar.
For the abundances of Na, the ratio of
ionized to neutral Na in the direction of $\zeta$~Oph has been adopted
for all other sight lines, although the validity of this assumption is unsettled. 

\begin{table*}
\caption{Comparison of interstellar abundances (first rows) with that of \LB\ stars (second/third rows). The Orion region corresponds to the sight lines to the stars HD\,35149, HD\,36486, HD\,36861 and HD\,37128, and the $\lambda$~Sco region corresponds to HD\,154368 and HD\,158926.}
\label{compare_tab}
\begin{tabular}{@{}r@{ }c@{ }c@{ }c@{ }c@{ }c@{ }c@{ }c@{ }c@{ }c@{ }c@{ }c}
\hline\hline
                   & C/C$_{\rm NIR}$   & O/O$_{\rm NIR}$  & Na        & Mg        &  Si       & S         & Ca        &  Ti       & Cr        & Mn     & Fe \\
\hline
ISM (HD 22586)     &                   &                  &           &           & $-$0.6(2) &           & $-$2.4(2) & $-$1.7(2) &           &        & $-$1.1(2) \\
HD 11413           &                   &                  &           &           & $-$0.2(2) &           & $-$0.9(3) & $-$1.4(2) &           &        & $-$1.5(2) \\
\hline
ISM (Orion region) &                   & $-$0.4(1)        & $-$2.1(3) &           &           &           &           &           & $-$1.7(3) & & \\
HD 31295           &                   & $-$0.5(1)/0.0(1) & $-$0.5(2) &           &           &           &           &           & $-$0.8(2) & & \\
\hline
ISM (HD 93521)     &                   &                  &           &   +0.0(4) & $-$0.4(2) &   +0.0(1) &           & $-$1.1(5) &           & $-$0.7(3) & $-$0.9(4) \\
HD  84123          &                   &                  &           & $-$1.0(2) & $-$1.0(2) & $-$0.6(1) &           & $-$1.0(1) &           & $-$1.1(3) & $-$1.2(2) \\
HD  84948 A        &                   &                  &           & $-$1.2(5) & $-$0.8(4) &           &           & $-$1.3(3) &           &           & $-$1.2(3) \\
HD  84948 B        &                   &                  &           & $-$1.0(4) & $-$0.6(5) &           &           & $-$0.5(4) &           &           & $-$1.0(2) \\
HD  101108         &                   &                  &           & $-$0.3(2) & $-$0.5(3) &           &           & $-$0.2(2) &           & $-$0.4(2) & $-$0.7(1) \\
\hline
ISM (3C 273)       &                   &                  &           &           & $-$0.7(2) &           &           &           &           & $-$0.7(1) & \\
HD 105759          &                   &                  &           &           &           &           &           &           &           & $-$1.0(5) & \\
HD 110411          &                   &                  &           &           & $<-$0.3   &           &           &           &           &           & \\
\hline
ISM (HD 157246)    &                   & $-$0.3(1)        &           &           &           &           &           &           &           & & \\
HD 168740          &                   & /0.0(1)          &           &           &           &           &           &           &           & & \\
\hline
ISM ($\lambda$~Sco region) & $-$0.4(2) & $-$0.5(1)        &           & $-$0.6(3) &           &           &           & $-$2.5(3) & $-$1.1(3) &           & $-$1.2(3) \\ 
HD 170680                  & /0.0(1)   & /0.0(1)          &           & $-$0.2(2) &           &           &           & $-$0.5(1) & $-$0.4(3) &           & $-$0.4(1) \\
\hline
ISM (HD 212571)    &                   &                  &           &           &           & $-$0.3(3) &           &           & $-$1.4(3) & & \\
HD 204041          &                   &                  &           &           &           &   +0.0(2) &           &           & $-$0.8(2) & & \\
\hline\hline
\end{tabular}
\end{table*}

\begin{figure}
  \resizebox{\hsize}{!}{\includegraphics{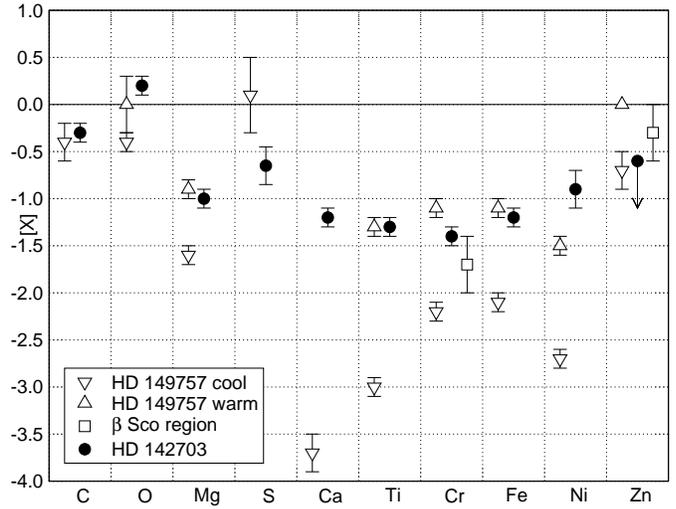}}
  \caption{Comparison of the interstellar abundances towards the regions of HD\,149757 
  and $\beta$~Sco
  (representing the sight lines towards HD\,141637, HD\,143018, HD\,144217 and
  HD\,147165) with that of the \LB\ star HD\,142703.}
  \label{hd149757}
\end{figure}

For S, the ISM abundance distribution is exactly the same as for the \LB\ stars.
On the other hand, the {\em mean} abundance of Zn is significantly
{\em lower} (by 0.6~dex) in \LB\ stars than in the ISM. 
Also, the mean abundances of Mg, Al, Si and Mn are slightly
lower (but note that only four ISM measurements are available for Al).
Furthermore, for all of these elements except Al the {\em lowest} abundances are lower 
in the \LB\ stars than in the ISM by 0.4 to 0.8~dex.

This does not seem to fit into the
accretion/diffusion theory, because the originally
normal element abundance in the stellar atmosphere should converge to the abundance
in the accreted (interstellar) matter. The ultimate atmospheric abundance
depends mainly on the accretion rate and the mass of the convection zone. 
In this picture the ISM abundances set a lower limit for the \LB\ star abundances, 
which is exceeded by the observations.
The situation is different for the other heavier elements,
where the mean ISM abundances lie well below the mean \LB\ abundances,
and also the lowest ISM abundances are lower than that of the \LB\ stars.
The most controversial abundances are that of Zn,
whose mean abundance and distribution is similar to that of S in the ISM.
The mean [Zn/Fe] ratio is +1.0 $\pm$0.3~dex in the ISM, whereas the mean ratio
of the same elements for \LB\ stars is +0.3 $\pm$0.4~dex. These values have
been derived including only sight lines or stars for which the abundances of
both elements are available. We obtain the same values when using only Fe~I
lines for the \LB\ stars, thus possible unconsidered Zn~I NLTE effects
seem to be negligible (see also \citetalias[ Sect.~4]{Heit:01a}).

The significance of these findings could be reduced 
by the fact that interstellar
depletions depend on cloud conditions \citep{Sava:96a,Spit:85}, 
but we have averaged over a large
range of densities (and temperatures). However, if we divide the ISM sample
into clouds with ``low'' ($\log{N(\rm H)} < 20.5$) and ``high'' 
($\log{N(\rm H)} > 20.5$) densities the average depletions of the
deficient elements are larger in the denser (and cooler) clouds by only 0.3~dex,
although the minimum abundances are on the average 
by 0.9~dex lower. Therefore, the problems discussed in the previous paragraph remain,
in particular if we regard the circumstellar environment to be represented
more closely by the less dense (and warmer) interstellar clouds.

\subsection{Individual sight lines}
\label{ISM_lb}
We compared also the abundance patterns of individual \LB\ stars with that of 
nearby ISM sight lines. 
The extension of the interstellar cloud in the direction of HD\,154368 is
given by \citet{Snow:96} 
as $l$=346\deg\dots 25\deg, $b$=$-$6\deg\dots +6\deg, r=125$\pm$25~pc. 
This region is located near the \LB\ star HD\,170680.
In analogy, sight lines having \LB\ stars with a 
($l,b$) distance of less than about 15\deg\
in their vicinity have been marked with an asterisk in Table~\ref{ISM_tab}.
This corresponds to linear distances of about ten to 60~pc, depending
on the distances of the \LB\ stars from the Sun.
The respective \LB\ stars are less distant than each of the ISM stars.
Only few common element abundances are available.
They are listed in Table~\ref{compare_tab}, where
upper rows and lower rows correspond to ISM sight lines and \LB\ stars,
respectively. One region with more elements available for a comparison 
is shown in Fig.~\ref{hd149757}.

Of particular interest is the region in the direction of HD\,93521, 
for which six elements can be compared with the stars HD\,84123,
HD\,84948 and HD\,101108. The ISM abundances of
the lighter elements Mg, Si and S are too high to explain the \LB\ abundances
by accretion of this (or similar) material. On the other hand, the abundances of the
heavy elements Ti, Mn and Fe are similar in this ISM region and the three \LB\ stars.
The averaged Cr and Zn abundances of four adjacent sight lines 
(HD\,141637, HD\,143018, HD\,144217 and HD\,147165 = ``$\beta$~Sco region''),
the abundance pattern of the most intensively studied $\zeta$~Oph sight line
and that of the \LB\ star HD\,142703 are displayed in Fig.~\ref{hd149757}.
All ISM abundances except for S are lower than in the
\LB\ star, when regarding the cool $\zeta$~Oph cloud. For the warm cloud,
the Cr and Zn abundances are also discrepant.
For the other regions we observe the same trends in the individual abundance
patterns as for the mean values.

In conclusion, the accretion scenario could explain the spectroscopic features of a
fraction of the \LB\ stars. For the other part, the observed ISM underabundances
are not low enough, which suggests that either a different mechanism is operating
or the circumstellar abundances differ significantly from that measured in the
near interstellar medium. It might also be possible that the local abundance
variations in the ISM are larger than assumed for the comparison.
Finally, there could be additional uncertainties in the stellar abundances
because of unaccounted-for physical processes in the atmospheres, like stratification
due to diffusion, or inadequate treatment of convection.

\section{Comparison with stars surrounded by circumstellar matter}
\label{post-AGB}

\begin{figure}
  \resizebox{\hsize}{!}{\includegraphics{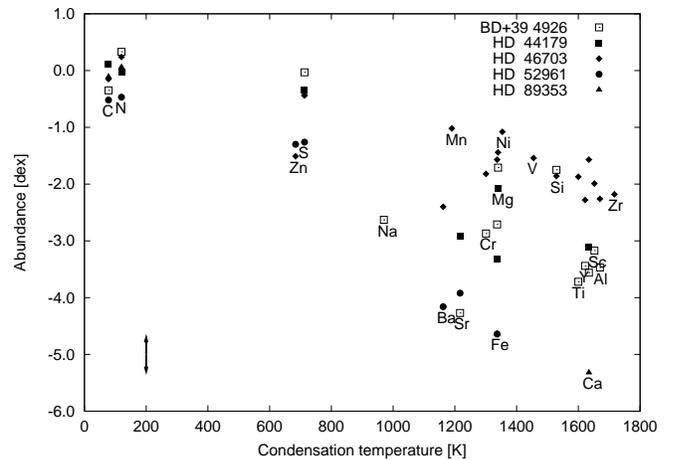}}
  \caption{Abundances of metal poor post-AGB stars versus condensation temperature. The arrow to the lower left represents the mean error of the abundances.}
  \label{cond_pAGB}
\end{figure}

\begin{figure}
  \resizebox{\hsize}{!}{\includegraphics{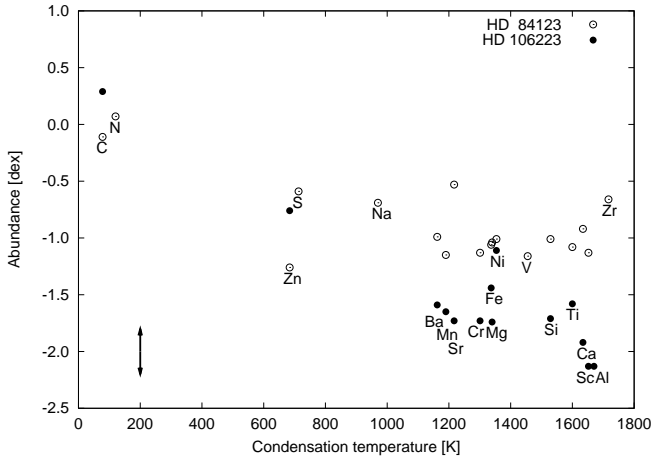}}
  \caption{Abundances of the \LB\ stars HD\,84123 and HD\,106223
           versus condensation temperature. The arrow to the lower left represents the mean error of the abundances.}
  \label{cond_LB}
\end{figure}

The {\em post-AGB} stars show evidence for being surrounded by circumstellar shells.
A small subgroup with unusually low metal abundances, but nearly solar abundances
of C, N, O and S has been identified \citep{Bond:91}. It has been proposed that they
have experienced accretion of the circumstellar gas depleted by dust fractionation
\citep{vanW:92}, which is supported by an enhanced abundance of Zn relative
to Fe in one of these stars.
Fig.~\ref{cond_pAGB} shows the correlation between
the abundances of the five metal poor post-AGB stars
\citep{Koda:73,Lamb:88,Luck:84,Bond:87,Wael:91,Wael:92,vanW:92}
and the condensation temperatures ($T_c$) of the elements \citep[taken from][]{Fegl:97}.
A similar but more shallow correlation (with one or two exceptional elements)
can be seen in HD\,106223
(Fig.~\ref{cond_LB}; note that the vertical scale is different from Fig.~\ref{cond_pAGB}) 
and six other \LB\ stars (HD\,319, HD\, 31295, HD\,75654, HD\,142703, HD\,204041
 and HD\,210111).
Other \LB\ stars show nearly solar abundances of elements with low condensation
temperature (C, N, in some cases also S and Na) and more or less constant
underabundances of the other elements like HD\,84123 in Fig.~\ref{cond_LB}.
We conclude that chemical separation could be responsible for the abundances
of some, but not all \LB\ stars, although the mechanism seems to be less
efficient than in the post-AGB stars.
For the comparison of the abundances of the two groups, it has to be taken
into account that the original mean abundance in post-AGB stars is low
\citep[$\approx -$0.7~dex, see][]{Heit:00} because of their age (population~{\sc ii}).
Furthermore, their C, N and O abundances have been enhanced during the AGB phase.

\begin{table}
\caption{Dust and gas around \LB\ stars. The abbreviations for the dust references are as follows: S1986 -- Sadakane and Nishida (1986), C1992 -- Cheng et al. (1992), K1994 -- King (1994), F1999 -- Fajardo-Acosta et al. (1999), P2000 -- Paunzen (2000). The references for gas detection are: 1 -- Holweger et al. (1999), 2 -- Hauck et al. (1998), 3 -- Hauck et al. (1995)}
\label{shell_tab}
\begin{tabular}{llp{2.5cm}ll}
\hline\hline
HD     & Dust?    & Ref.                & Gas? & Ref. \\
\hline
319    & ?        & --                  & no   & 1 \\
11413  & probably & F1999               & yes  & 1 \\
30422  & ?        & --                  & no   & 1 \\
31295  & yes      & K1994,\,S1986       & no   & 1, 2 \\
109738 & no       & K1994               & ?    & -- \\
110411 & yes      & F1999,\,C1992       & no   & 1, 2 \\
111604 & ?        & --                  & no   & 2 \\
125162 & yes      & P2000,\,K1994, S1986 & no   & 2 \\
142703 & probably & P2000               & no   & 1 \\
154153 & no       & K1994               & ?    & -- \\
183324 & ?        & --                  & no   & 1 \\
192640 & no/yes   & P2000/F1999         & no   & 3 \\
193256 & ?        & --                  & yes  & 1 \\
193281 & ?        & --                  & no   & 1 \\
198160/1 & ?      & --                  & yes  & 1 \\
204041 & probably & P2000               & no   & 1 \\
210111 & ?        & --                  & no   & 1 \\
221756 & no       & P2000               & no   & 2 \\
\hline\hline
\end{tabular}
\end{table}

{\em ``Vega-like''} stars exhibit IR flux excesses which indicate the presence of a 
dust envelope. Many members of this class have been found through searches
of the IRAS catalog \citep[ and references therein]{Mann:98}.
Three \LB\ stars are amongst them: $\lambda$~Boo, $\pi^1$~Ori and $\rho$~Vir.
Table~\ref{shell_tab} summarizes the results of 
dedicated searches for IR excesses in \LB\ stars
including data obtained with the ISO satellite.
As the dust detection rate is very small, the existence of shells
around most \LB\ stars, as required by the accretion/diffusion
hypothesis, has to be questioned. Otherwise, the excess IR emission 
might be too faint to be detected with today's IR photometric devices.
\citet{Holw:99} investigated high resolution spectra of normal 
and ``dusty'' A stars and \LB\ stars (Table~\ref{shell_tab}).
They concluded that signatures of circumstellar gas are not as common 
as that of circumstellar dust in normal A stars 
and that either the two kinds of circumstellar
matter rarely coexist around a star (they could appear at different evolutionary 
stages) or that both components cannot be detected at the same time. 

Abundances are only available for seven ``dusty'' A type main sequence stars
\citep{Dunk:97,Holw:97}, apart from Vega \citep{Adel:90b,Ilij:98}.
Two of them ($\beta$~Pic and $\gamma$~Oph) additionally show evidence for 
circumstellar gas shells in their spectra.
No abundances are available for any other spectroscopically classified 
``A-shell'' stars.
The abundances of Vega lie in the range of the \LB\ star abundances,
except for Al, V and Zr, which are more abundant than in any of the \LB\
stars. This abundance pattern provides only weak evidence against a 
classification of Vega as a \LB\ star. 
But a recent analysis of IUE spectra (E. Solano, private communication),
shows that Vega has to be excluded from the \LB\ group, because
the C/(Al,Si,Ca) equivalent width ratios are by far smaller than the
limiting criteria defined by \citet{Sola:98,Sola:99}.
However, Vega is the most metal deficient of the analysed Vega-like stars, 
with only one star (HD\,169142) reaching similar underabundances for Mg and Si. 
All other element abundances correspond to that of normal stars \citepalias{Heit:01a}.
These results show that presence of circumstellar matter around A-type
main sequence stars is not necessarily related to abundance anomalies as observed
in \LB\ stars.


\section{Conclusions}

\begin{table}
\caption{Summary of the search for signatures of accretion. $\sigma$ \dots\ cool stars with small (+) or large ($-$) $\sigma$ from Table~\ref{par_all_tab}, hot stars with [Ca]=[Ti] ({\sf x}); I \dots\ all abundances higher than in ISM (+), some abundances lower than in ISM ($-$); C \dots\ smooth (+) or discontinuous ($-$) relation of abundances and $T_c$.}
\label{comb}
\begin{tabular}{rccc|rccc}
\hline\hline
HD & $\sigma$ & I & C & HD & $\sigma$ & I & C \\
\hline
    319   &  +  &     &  +   &   125162   & $-$ &     & $-$ \\
  11413   & $-$ &  +  & $-$  &   142703   &  +  & $-$ &  +  \\
  15165   & $-$ &     & $-$  &   168740   &  +  &     &     \\
  31295 &{\sf x}&  +  &  +   &   170680   &     &  +  &     \\
  74873 &{\sf x}&     & $-$  &   183324 &{\sf x}&     & $-$ \\
  75654   &     &     &  +   &   192640   &  +  &     & $-$ \\
  84123   &  +  & $-$ & $-$  &   193256   & $-$ &     & $-$ \\
 84948A/B &  +  & $-$ &      &   198160/1 & $-$ &     &     \\
 101108   &  +  &  +  & $-$  &   204041   & $-$ &  +  &  +  \\
 106223   &  +  &     &  +   &   210111   & $-$ &     &  +  \\
 107233   &  +  &     &      &   221756 &{\sf x}&     &     \\
 110411 &{\sf x}&     & $-$  & & & & \\ 
\hline\hline
\end{tabular}
\end{table}

The \LB\ star abundances were examined with regard to correlations to the
stellar parameters of this group, in particular the effective temperature.
It was found that for some elements
(C, Na, Mg, Si, Ca, Sc, Cr, Fe, Sr) the abundances are weakly correlated 
with \Teff, \logg, the age, the pulsational period or \Vsini.
The scatter of heavy element abundances in individual stars does not depend on \Teff.
These findings are inconclusive with regards to testing the accretion/diffusion theory.
Because of the lack of calculations
for more than three elements and different atmospheric parameters, the uncertainties related to the treatment of convection and
the calculation of the radiative acceleration, and the free parameters 
(mainly the accretion rate and the abundance spectrum in the accreted material),
we consider the theoretical models to be rather simple and incomplete. 

The chemical composition of the \LB\ stars has been compared to that of the
interstellar medium (ISM).
The {\em mean} abundances of some elements (Mg, Si, Mn, Zn) are slightly lower 
in the \LB\ stars than in the ISM (by 0.2 to 0.6~dex), and the {\em lowest}
abundances found in \LB\ stars for these elements are lower than the lowest
ISM abundances by 0.4 to 0.8~dex. Similar deviations have been found for
only half of the single stars which can be compared to nearby sight lines.
Within an accretion/diffusion scenario, the abundances of the accreted elements 
would be expected to be greater than in the ISM.
 
The \LB\ abundance pattern has also been compared to
that of stars with circumstellar material (post-AGB, Vega-like and A-shell stars).
Similar relations of abundances with condensation temperatures 
suggest that the same physical processes lead to the chemical
compositions of some \LB\ stars and the metal poor post-AGB stars
although theoretical calculations for the latter group do not exist.
More observations are clearly needed to confirm this hypothesis.
On the other hand, the lack of metal deficiency in dusty
stars with atmospheric parameters similar to \LB\ stars questions the connection
of circumstellar dust with the \LB\ phenomenon, although the comparison 
is based on a very small sample of Vega-like stars.

From the currently available abundance data we conclude that the stars HD\,319,
HD\,31295 and HD\,106223 could well have experienced accretion of circumstellar gas
(see Table~\ref{comb}), which however has not been detected in their spectra.
For the other stars, further examination and spectral data are required.


\begin{acknowledgements}
This research was carried out within the working group {\em
Asteroseismology--AMS}, supported by the Fonds zur F\"orderung der 
wissenschaftlichen Forschung (project {\sl S\,7303-AST}).
We want to thank E. Solano for providing the information on the IUE spectra
of Vega. 
Thanks go to the referee, K.A. Venn, whose comments have helped to greatly
improve the paper.
Use was made of the Simbad database, operated at CDS, Strasbourg, France.
\end{acknowledgements}

\bibliographystyle{aa}
\bibliography{LB}

\end{document}